\documentclass[fleqn,12pt,twoside]{article}
\usepackage{espcrc1}
\usepackage{epsfig}

\newcommand{\AmS}{{\protect\the\textfont2
  A\kern-.1667em\lower.5ex\hbox{M}\kern-.125emS}}

\title{ Two Particle Azimuthal Correlation Measurements in PHENIX }

\author{ N. N. Ajitanand\address[SBChem]{Chem. Dept. Stony Brook University, 
New York, NY 11794-3400, USA} for the PHENIX Collaboration\thanks{for the full 
PHENIX Collaboration author list and acknowledgments, see Appendix "Collaborations" of this volume. }
}

\begin{document}

\vskip -2cm

\maketitle


\begin{abstract}
             {\small Two particle azimuthal correlation functions are presented 
for charged hadrons produced in Au~+~Au collisions at RHIC ($\sqrt{s_{_{NN}}}=200$~GeV).
The correlation functions indicate sizeable asymmetries and anisotropies. 
The trend of the asymmetries is compatible with the presence of 
emission patterns associated with mini-jets. 
The magnitude and the trend of the differential anisotropies 
$v_2(p_T)$ and $v_2(N_{Part})$, provide important model constraints.
}
\end{abstract}

\vskip -1.0cm 
\section{Introduction}

Differential measurement of azimuthal ($\Delta\phi$) correlation functions for 
charged hadrons is arguably one of the most important experimental probes for  
high energy-density nuclear matter created  at the Brookhaven Relativistic Heavy Ion 
Collider (RHIC). The anisotropy as well as the asymmetry of such correlation functions 
are predicted to; (i)~serve as a ``barometer" for collective transverse flow and 
hence a probe for the equation of state (EOS)~\cite{Teaney_hydro:2001,Kolb_hydro:2001},
(ii)~provide important constraints for the density and effective energy loss of 
partons~\cite{gvw_jq:2001,molnar_op:2001,muller_jq:2002}, and (iii)~provide invaluable 
information on nuclear saturation scales~\cite{kovchegov_mj:2002}. 
Full exploitation of these correlation functions require a good understanding 
of the mechanistic origin of two-particle azimuthal correlations at RHIC energies.
Detailed systematic measurements of such correlations constitute important steps in this 
endeavor.  

\vskip -.75cm 
\section{Analysis and Results}

 	The colliding Au beams ($\sqrt{s_{_{NN}}}=200$~GeV) used in these 
measurements have been provided by the RHIC. 
Charged particles from mimum-bias triggered events were 
detected in the fully instrumented east and west central arms of 
PHENIX~\cite{nagle_qm2002}. Each of these arms subtends 90$^{o}$ 
in azimuth $\phi$, and $\pm 0.35$ units of pseudo-rapidity $\eta$. 
The axial magnetic field of PHENIX (0.5 T) allowed for the tracking of 
particles with $\rm{p} > 0.2$ GeV/c 
(${\delta {\rm{p}}/{\rm{p}} \simeq 1\% }$) in the 
fiducial volume of both arms. Good track quality was ensured via matching 
and veto cuts to outer detectors in 
each arm. The Zero Degree Calorimeters (ZDC), were used in conjunction with the 
Beam-Beam Counters (BBC), to provide off-line selections of a wide range of 
centralities expressed as a fraction of the total interaction cross section. 

Two-particle azimuthal correlation functions  were constructed 
via the ratio of two distributions~\cite{lacey_qm2001};
 $C(\Delta\phi) = {N_{cor}(\Delta\phi)/{N_{mix}(\Delta\phi)}}$, 
%
where $N_{cor}(\Delta\phi)$ is the observed $\Delta\phi$ distribution
for charged particle pairs selected from the same event,  
and $N_{mix}(\Delta\phi)$
is the $\Delta\phi$ distribution for particle pairs selected from
mixed events. Mixed events were obtained by randomly selecting each member 
of a particle pair from different events having similar centrality 
and vertex position. Two correlation functions were obtained for each 
$p_T$-range of interest. In the first, charged hadron pairs were formed by 
selecting both particles from a reference range \mbox{$p_{T_{Ref}}$}, which excludes the 
$p_T$ range of interest (i.e. a reference correlation~\cite{lacey_qm2001}).
In the second, hadron pairs were formed by selecting one member from 
the $p_T$ range of interest, and the other from the reference  
range $p_{T_{Ref}}$ (i.e. an assorted-$p_T$ correlation~\cite{lacey_qm2001}).
 
\vskip -.75cm
\begin{figure}[h]
\begin{minipage}[htb]{15.5cm}
\epsfig{file=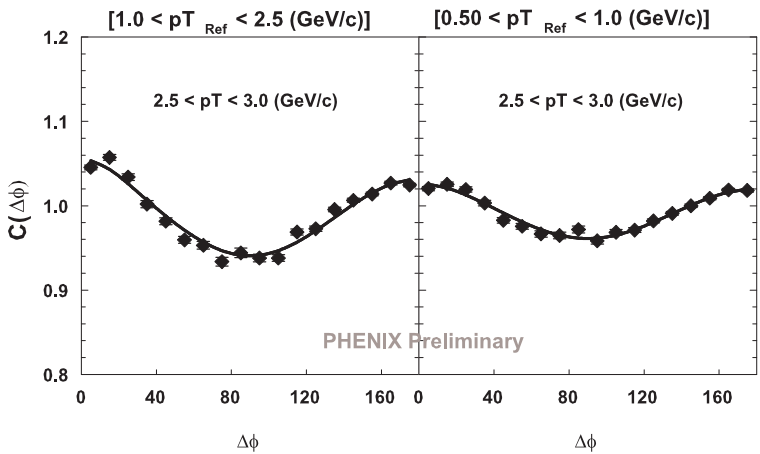,width=10.5cm}
\vskip -.75cm
\caption{\small  
	Assorted-$p_T$ correlation functions for charged hadrons of $2.5 < p_T < 3.0$ 
GeV/c. Correlation functions are shown for \mbox{$1.0 < p_{T_{Ref}} < 2.5$} GeV/c (left panel) 
and for \mbox{$0.5 < p_{T_{Ref}} < 1.0$} GeV/c (right panel). The solid line represents a 
fit to the correlation function, see text. 
}
\label{fig:cor_fun}
\end{minipage}
\end{figure}
\vskip -.75cm
        Figure~\ref{fig:cor_fun} shows representative differential 
assorted-$p_T$ correlation functions for an event centrality cut of 20-40\%.
The hadron pairs were formed by selecting one member from the range 
\mbox{$2.5 < p_T < 3.0$} GeV/c and the other from
a non-overlapping reference range $p_{T_{Ref}}$. An additional requirement
that particle pairs detected in the same PHENIX arm have the same charge 
(++ or -\,-) was also imposed to reduce [but not exclude] non-flow correlations. 
The right panel of Fig.~\ref{fig:cor_fun} indicates an essentially 
symmetric correlation function for \mbox{$0.5 < p_{T_{Ref}} < 1.0$} GeV/c.
By contrast the correlation function for \mbox{$1.0 < p_{T_{Ref}} < 2.5$} GeV/c (left panel) 
shows an asymmetry at small $\Delta\phi$. 
This asymmetry grows if narrow pseudo-rapidity cuts ($\Delta\eta <0.35$) are 
imposed or if particle pairs are required to have dissimilar charge (+\,-). 
These trends are compatible with 
the effects of jet fragmentation~\cite{mchiu_qm2002}.

\begin{figure}[h]
\begin{minipage}[htb]{15.5cm}
\epsfig{file=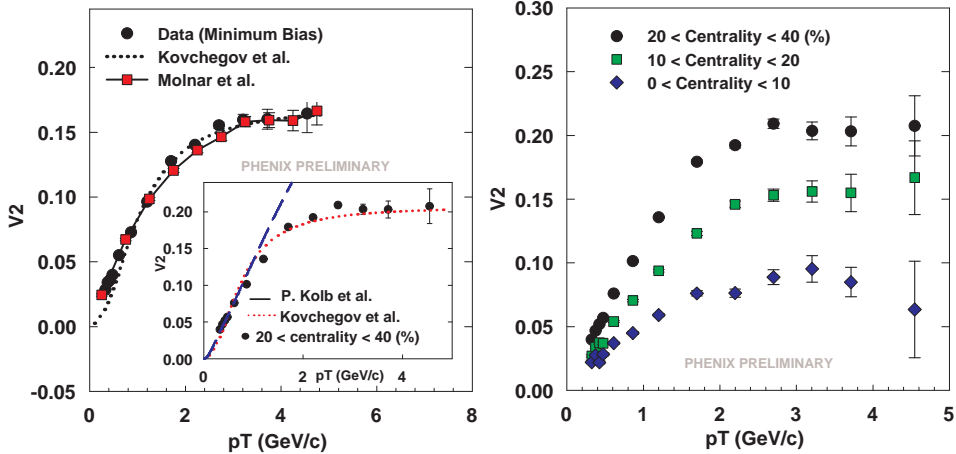,width=13.0cm}
\vskip -.75cm
\caption{\small  $v_2$ vs. $p_T$ for minimum bias events (left panel) and for
several different centralities as indicated (right panel). Model comparisons
\cite{molnar_op:2001,kovchegov_mj:2002}, 
are indicated for the minimum bias data. The inset shows model comparisons \cite{Kolb_hydro:2001,kovchegov_mj:2002} to data for the centrality cut of 20-40\%.
}
\label{fig:pt_dep}
\end{minipage}
\end{figure}

\begin{figure}[h]
%
\begin{minipage}[htb]{15.5cm}
\epsfig{file=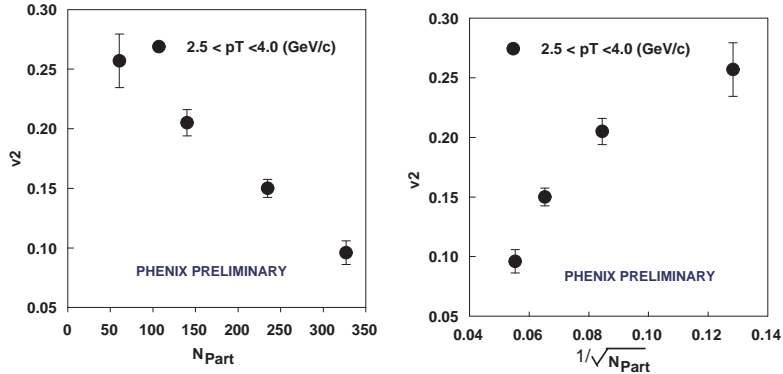,width=10.5cm}
\vskip -.75cm
\caption{\small  $v_2$ vs. $N_{Part}$ (left panel), and 
 $\sqrt{N_{Part}}$, (right panel) for charged hadrons of $2.5 < p_T < 4.0$.
Error bars indicate statistical errors only.}
\label{fig:cent_dep}
\end{minipage}
\end{figure}

	The anisotropy and asymmetry of the generated correlation functions have been
characterized via fits. A good representation of the data for the 
reference-$p_T$ and the assorted-$p_T$ correlation functions is obtained with the fit functions
%
\[
C(\Delta \phi)= \lambda \cdot exp(-0.5(\Delta \phi/\sigma)^2) + 
a_1(1+ 2 v_{2_f}^2 cos(2\Delta \phi))
\]
%
\[
C(\Delta \phi)= \lambda \cdot exp(-0.5(\Delta \phi/\sigma)^2) + 
a_1(1+ 2v_{2_a} cos(2\Delta \phi))
\]
%
respectively. Here the Gaussian term and 
the $cos(2\Delta \phi)$ term is used to characterize the asymmetry 
[at small $\Delta \phi$] and the anisotropy respectively. 
The anisotropy parameter for the $p_T$ range of interest $v_2'$, is 
given by $v_2'$ = $v_{2_a}$/$v_{2_f}$ where $v_{2_a}$ 
and $v_{2_f}$ are the anisotropies 
extracted from the assorted and the reference correlations respectively. 
The value of $v_2'$ can be compared to the second Fourier 
coefficient $v_2$, commonly used to quantify the anisotropy with respect to 
the reaction plane~\cite{shinichi_qm2002}.
 Differential  values of $v_2'$ (hereafter referred to as $v_2$) are summarized 
in Figs.~\ref{fig:pt_dep} and \ref{fig:cent_dep}.  
\vskip -.75cm
\vskip -.75cm
%
Figure~\ref{fig:pt_dep} shows differential anisotropies $v_2(p_T)$, which increase with 
$p_T$ followed by an apparent saturation for 
$p_T~\mathbin{\lower.3ex\hbox{$\buildrel>\over
{\smash{\scriptstyle\sim}\vphantom{_x}}$}}$~2.5 GeV/c. This trend is similar
for all centralities shown and is without doubt related to the collision dynamics.
Hydrodynamic models~\cite{Teaney_hydro:2001,Kolb_hydro:2001} provide a good description 
of the data for $p_T$ up to $\sim 1.5$ GeV/c. However, these models do not predict the 
observed saturation of $v_2(p_T)$ 
(see inset). The lack of saturation indicates that the model assumption that 
local equilibrium can be maintained until a sudden freeze-out hypersurface is reached, 
is invalid outside some finite domain of phase space. Results obtained from the 
jet dominated model HIJING, also show poor agreement with the data  at high $p_T$.
Models incorporating only strong jet quenching~\cite{muller_jq:2002} or 
jet quenching acting in concert with hydrodynamic expansion~\cite{gvw_jq:2001} do
not provide a good description of the high-$p_T$ data as well. By contrast, results 
from a recent implementation of a covariant transport theory~\cite{molnar_op:2001} gives a good 
representation of the data if either extremely high initial gluon density or very 
large parton-parton scattering cross sections are employed in the calculations (see left panel
of Fig.~\ref{fig:pt_dep}). 
Results from a saturation model~\cite{kovchegov_mj:2002} can also account 
for the observed anisotropies (see left panel and inset of Fig.~\ref{fig:pt_dep}).
 
	The mechanistic origin of two-particle correlations in the transport and 
saturation models are very different. Consequently, additional constraints are required to 
facilitate a distinction between them. One such constraint is the centrality dependence 
of $v_2$. The saturation model predicts that $v_2$ should scale 
as $1/\sqrt{N_{Part}}$~\cite{kovchegov_mj:2002}, where $N_{Part}$ represents
the number of participants in the collision. For large opacities 
(high initial gluon density or very large parton-parton scattering cross sections) 
one expects the transport model to exhibit essentially $N_{Part}$ scaling.
	Fig.~\ref{fig:cent_dep} show initial tests for $1/\sqrt{N_{Part}}$ scaling ( right panel), and 
$N_{Part}$ scaling (left panel) for charged hadrons of $2.5 < p_T < 4.0$ GeV/c. The data 
indicates better overall agreement with $N_{Part}$ scaling for the centrality range 
presented but does not exclude $1/\sqrt{N_{Part}}$ for relatively central collisions. 

\vskip -1.0cm 
\section{Summary}

	In summary, two-particle correlation 
measurements indicate sizeable anisotropies and asymmetries.
The trend of the asymmetries are compatible with the asymmetric emission pattern of 
mini-jets. The anisotropy ($v_2$) increases with decreasing centrality and increases 
with $p_T$ up to $p_T\sim 2.5$ GeV/c. For higher $p_T$'s the anisotropy saturates for 
each centrality cut. This important feature is not reproduced by hydrodynamic models 
but is well reproduced by the covariant transport theory 
of Molnar {\it et al}.~\cite{molnar_op:2001}, and the saturation model of 
Kovchegov {\it et al}.~\cite{kovchegov_mj:2002}. More detailed measurements of the 
centrality dependence of $v_2$ may allow one to distinguish between the two 
very different mechanisms for two-particle correlations, implied by the latter models.

\end{document}